\documentclass[%
 reprint,
 amsmath,amssymb,
 aps,
]{revtex4-1}

\usepackage{graphicx}
\usepackage{dcolumn}
\usepackage{bm}
\newcommand{\z}{\overline{z}}
\newcommand{\E}{\overline{E}}
\newcommand{\Ee}{\tilde{E}}
\usepackage{color}

\begin{document}

\preprint{APS/123-QED}

\title{Forecasting failure locations in two-dimensional disordered lattices}

\author{Estelle Berthier}
  \email{ehberthi@ncsu.edu}
\affiliation{%
 Department of Physics, North Carolina State University, Box 8202, Raleigh, NC 27695, USA.
}

\author{Mason A. Porter}
\affiliation{
Department of Mathematics, University of California, Los Angeles, 520 Portola Plaza, Los Angeles, California 90095, USA.
}

\author{Karen E. Daniels}
\affiliation{
Department of Physics, North Carolina State University, Box 8202, Raleigh, NC 27695, USA.
}

\medskip

\date{\today}

\begin{abstract}
Forecasting fracture locations in a progressively failing disordered structure is of paramount importance when considering structural materials. We explore this issue for gradual deterioration via beam breakage of two-dimensional disordered lattices, which we represent as networks, for various values of mean degree. We study experimental samples with geometric structures that we construct based on observed contact networks in 2D granular media. We calculate geodesic edge betweenness centrality, which helps quantify which edges are on many shortest paths in a network, to forecast the failure locations. We demonstrate for the tested samples that, for a variety of failure behaviors, failures occur predominantly at locations that have larger geodesic edge betweenness values than the mean one in the structure. Because only a small fraction of edges have values above the mean, this is a relevant diagnostic to assess failure locations. Our results demonstrate that one can consider only specific parts of a system as likely failure locations and that, with reasonable success, one can assess possible failure locations of a structure without needing to study its detailed energetic states.
\end{abstract}

\maketitle

\section{Introduction} \label{sec:intro}
Cellular foams \cite{gibson_mechanics_1982}, 
semiflexible fiber and polymer networks \cite{broedersz_modeling_2014}, and many recently-developed mechanical metamaterials \cite{bertoldi_flexible_2017, goodrich_principle_2015, rocks_designing_2017} all belong, in idealized form, to a general class of disordered lattices. Such lattices can range in size from microscopic scaffolds for biological tissue growth \cite{janmey2008} to modern architectural structures \cite{knippers2016}. In each case, one can further idealize the material or structure as a mathematical network of connections between slender beams that intersect at various points within the material. From an engineering perspective, such materials are promising because of their light weights and their tunable, designable properties: a Poisson ratio from the auxetic \cite{hanifpour_mechanics_2018, reid_auxetic_2018,goodrich_principle_2015} to the incompressible limits \cite{goodrich_principle_2015}, a targeted local response to a remote perturbation \cite{rocks_designing_2017}, or the ability to change shape \cite{bertoldi_flexible_2017}. A disadvantage of these materials is that those that are constructed from stiff materials can degrade progressively through successive abrupt failures of the beams during loading \cite{driscoll_role_2016,hanifpour_mechanics_2018,zhang_fracturing_2018}.  To design optimized structures and safely use them for structural applications, it is necessary to assess the most likely locations of  fracture. Such predictive understanding would further enable the design of a material to fail in a prescribed way.

Fracture experiments have been conducted previously on printed, disordered auxetic materials \cite{hanifpour_mechanics_2018} and laser-cut, disordered honeycomb two-dimensional (2D) lattices \cite{driscoll_role_2016}. In these studies, very different fracture behaviors (ductile versus brittle) have been obtained by changing the loading direction \cite{hanifpour_mechanics_2018} or tuning the rigidity \cite{driscoll_role_2016}. In the latter study, a clear change arose in the spatial organization of fractures: they either can be dispersed throughout a system or be localized in the form of a narrow crack. Therefore, although some tunable parameters for controlling failure behavior have been identified, what determines these particular failure locations remains an open question. According to Griffith theory \cite{griffith_vi._1921}, damage in brittle materials focuses at the tip of a crack. However, factors such as material disorder \cite{roux_rupture_1988,alava_role_2008, shekhawat_damage_2013,curtin_brittle_1990,kahng_electrical_1988}, material rigidity \cite{driscoll_role_2016}, and the connectivity (specifically, mean degree) of networks \cite{driscoll_role_2016,zhang_fiber_2017} can affect the spatial organization of damage. As one tunes each of these factors, one can make failures spread throughout a system (diffuse damage), rather than forming a narrow crack (localized damage).

Zhang \textit{et al.} \cite{zhang_fracturing_2018} showed recently that failures can also be delocalized in topological Maxwell lattices (in which freely-rotating joints that are linked by rigid struts are on the verge of mechanical instability) \cite{mao2018}.
They performed numerical experiments on the tensile fracture of deformed square and kagome lattices, demonstrating that stress and fracture concentrate on self-stress domain walls, even in the presence of damage that is introduced elsewhere in the system. In another recent paper, Tordesillas \textit{et al.} \cite{tordesillas_interdependent_2018} studied damage locations in discrete-element simulations of concrete samples under uniaxial tension. From a network-flow analysis of the contact-network topology and contact capacities of a specimen, the authors determined the location of the principal interacting macrocracks. In their samples, they observed that secondary macrocracks develop in the pre-failure regime after damage occurs elsewhere, but before the formation of a dominant macrocrack, which sets the ultimate failure pattern of a sample.

\begin{figure}
	\center
	\includegraphics[width=\linewidth]{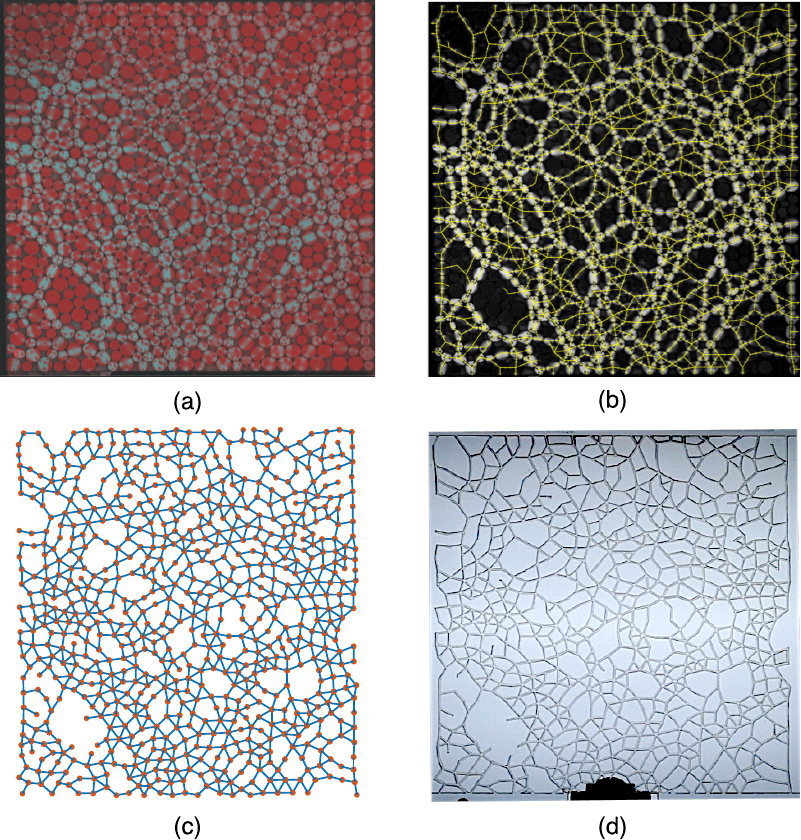} 
	\caption{
	(a) Force chains (cyan) recorded in a two-dimensional assembly of frictional photoelastic disks (red), which we image via a circular polariscope \cite{daniels_photoelastic_2017}. Brighter particles carry stronger forces. 
	(b) Contact network (yellow), which we extract using an open-source photoelastic solver \cite{jekollmer_pegs:_2018}, overlaid on the reconstructed ``pseudo-image'' \cite{daniels_photoelastic_2017}. 
	(c) Network representation in which each particle center is a node (orange dots) and each load-bearing contact is an edge (blue lines) \cite{papadopoulos_network_2018}. 
	(d) Corresponding physical sample that we laser-cut from an acrylic sheet, with the edges represented by beams that intersect at crosslinks (which correspond to the nodes in the network). } 
	\label{fig:method}
\end{figure}

In the present paper, we investigate where damage occurs in disordered lattices that consist of identical-width beams, with a network topology specified by the contacts measured from a real, quasi-2D granular packing (see Fig.~\ref{fig:method}).
We identify a common property, a large value of {\it geodesic edge betweenness centrality} \cite{girvan_community_2002},  that is shared by the failure locations of progressive damage events of our tested samples.
Even without modeling the physical interactions between nodes, this property provides a diagnostic for identifying likely failure locations. Such an indicator would permit assessing these locations in
a structure, without studying its detailed energetic or stress states.
Ultimately, the choice of a granular-inspired geometry for the  disordered lattice will provide a route toward generalizing these studies across inherently different systems, which are linked by their network topology.

For each network, we laser-cut an acrylic sheet using a contact network that matches the one observed in a packing, and we then test its behavior under compressive or tensile loading. 
Because the set of contacts in a packing forms a network that is embedded in a plane, a lattice does as well. Such a lattice network consists of edges (representing the beams of the lattice) that intersect at nodes, which occur at the crosslinks of the lattice. Conceptually similar structures occur for streets and intersections in the study of road networks \cite{crucitti_centrality_2006,lee2014}, 
connections between internet routers, plant veins \cite{katifori_quantifying_2012}, fungi \cite{lee2017}, and many other spatial systems \cite{barthelemy_spatial_2011, barthelemy_morphogenesis_2018}. 

Network analysis provides useful approaches --- including measures, algorithms, and theory --- for characterizing  complex spatial systems at multiple scales, ranging from local features to mesoscale and macroscale ones, and examining how they evolve  \cite{newman2018,papadopoulos_network_2018}. As discussed by Smart \textit{et al.} \cite{smart_granular_2008}, it is appealing to investigate what insights network analysis and associated topics (e.g., graph theory and algebraic topology) can yield on novel physical systems, especially in comparison to traditional approaches. For example, this perspective was adopted by Tordesillas \textit{et al.} \cite{tordesillas_interdependent_2018} to study quasi-brittle failure using network flow. Such approaches have also been useful for the study of mesoscale structures, such as dense communities of nodes, in granular systems \cite{bassett_influence_2012}. Therefore, network analysis appears to be a promising route to identify common analytical tools that are capable of relating failure behaviors across a variety of disordered systems.

One important approach in network analysis is the calculation of ``centrality'' measures to ascertain the most important nodes, edges, and other subgraphs in a network \cite{newman2018,faust1994}. One particularly popular type of centrality, known as betweenness centrality, measures whether one or more parts of a network lie on many short paths; it has been employed to characterize the importances of nodes \cite{freeman_set_1977}, edges \cite{girvan_community_2002}, and other subgraphs. The most common type of betweenness centrality uses geodesic (i.e., strictly shortest) paths. 

Recently, in a study of granular materials, Kollmer \textit{et al.} \cite{kollmer2019betweenness} showed that there is a positive correlation between the geodesic betweenness centrality value of a node and the pressure on the corresponding particle. Previously, Smart \textit{et al.} \cite{smart_effects_2007} reported that edges with large geodesic betweenness centrality exert a strong influence on heat transport in granular media. Inspired by these investigations, we selected from among the variety of network measures \cite{newman2018,barthelemy_morphogenesis_2018} and focus on calculating geodesic edge betweenness centrality (GEBC). See \eqref{ebc-def} in Materials and Methods for its definition.

As was reported in Berthier \textit{et al.} \cite{berthier_rigidity_2018}, one can control the compressive and tensile failure behaviors of a disordered lattice by tuning the mean degree of its associated network. This control parameter provides a way to create systems with a variety of failure behaviors, ranging from ductile-like to brittle-like failure. In the present paper, we show for samples across the spectrum from brittle-like to ductile-like failure (see ``Mechanical testing protocol'' in Materials and Methods) that individual beam failures occur predominantly on edges with GEBC values that are above the mean of the network. From this result, we conclude that GEBC is a useful diagnostic for forecasting possible failure locations in our contact networks. We demonstrate the ability of an GEBC-based test, which consists of comparing the geodesic edge betweenness centrality value of an edge to a threshold value, to discriminate between beams that fail and those that remain intact. 
This finding, together with the work of \cite{smart_effects_2007}, suggests that
betweenness centrality is a useful measure for capturing essential physical properties in disordered systems. Our study also confirms that tools from network analysis give a promising paradigm for the study of fracture.

The effectiveness of GEBC, which depends on network topology rather than on specifying mechanical interactions, is unexpected. This motivates a deeper analysis to determine which behaviors do not depend primarily on the detailed physical properties of a system, but instead depend on its geometry (and associated network topology). 
Our use of unweighted networks focuses our investigation on network topology, and we compare results for both a measure (specifically, GEBC) that ignores the physics   
and a well-known scalar electrical analogy of elasticity known as a random-fuse network (RFN) model \cite{de_arcangelis_random_1985,kahng_electrical_1988}. 
The RFN model identifies the most stressed beams as the edges with the largest currents, as determined by solving Kirchhoff's laws, for a given voltage drop across the boundaries.
We show that the RFN model, even with its incorporation of physical considerations, does not markedly improve performance over GEBC. 
This indicates that one can capture essential features of the lattice failure behavior by geometric (rather than physical) considerations.


\section{Results}\label{results}

\subsection{Spatial heterogeneity and changes with applied strain of geodesic edge betweenness centrality}

\begin{figure}
\center
\includegraphics[width=\linewidth]{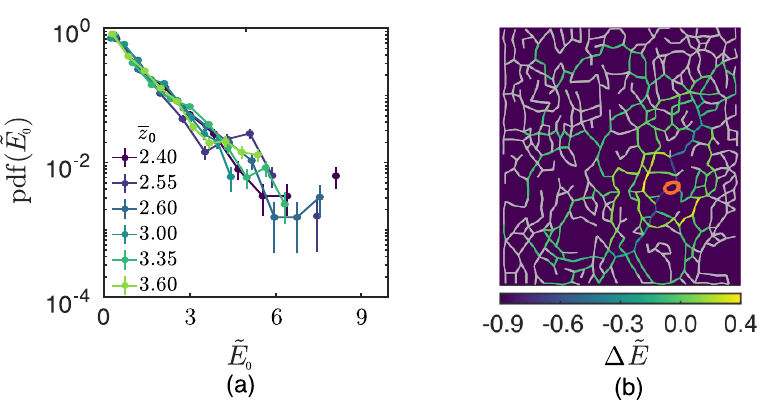} 
\caption{Characterization of geodesic edge betweenness (re)distribution: 
(a) Probability density function (PDF) of the initial geodesic edge betweenness centrality $\Ee_0$ for the different initial networks. We show the PDFs for several values of mean degree.
(b) Changes in geodesic edge betweenness centralities $\Ee$ after a failure event that occurs at the red ellipse at a compressive strain of $\varepsilon \approx 1.98\%$ on a network with mean degree $\overline{z}_0 = 2.40$. The values of the lavender edges change by less than $10^{-2}$.
}
\label{fig:EBC}
\end{figure}

We examine the ability of geodesic edge betweenness centrality $\Ee$ (see ~\eqref{ebc-def} in Materials and Methods) to forecast the specific locations at which our samples fail. For each initial (and subsequently altered)  network, we find that geodesic edge betweenness takes a broad range of values across the network. In Fig.~\ref{fig:EBC}a, we show the probability density function of the initial geodesic edge betweenness $\Ee_0$ for each initial network at each value of the mean degree $\z_0$. To facilitate notation, we use the subscript $0$ to designate our initial networks and the quantities that we measure and compute with them. 
In all cases, the distribution of values is approximately exponential, and it is largely independent of $\z_0$. Because each failure event (with associated edge removals) results in a new set of shortest paths, we obtain a new distribution of geodesic edge betweennesses for each altered network. Just as stress redistributes after damage \cite{alava_statistical_2006,zapperi_plasticity_1997,hidalgo_fracture_2002,berthier_damage_2017}, geodesic edge betweenness (due to its nonlocal nature) also redistributes in a system. In Fig.~\ref{fig:EBC}b, we show a characteristic example of redistribution after a failure event. The redistributions are system-wide: some edges are ``reloaded'', becoming more important with respect to the others (i.e., $\Ee_{s+1}>\Ee_{s}$, when going from strain step $s$ to strain step $s+1$),  
others are ``unloaded'' (i.e., $\Ee_{s+1}<\Ee_{s}$), and some edges (in lavender) have the same (or almost the same) value. By contrast, removal of unimportant edges (i.e., those with small values of geodesic edge betweenness) results in small (in amplitude) changes.

\begin{figure*}
\centering
\includegraphics[width=\linewidth]{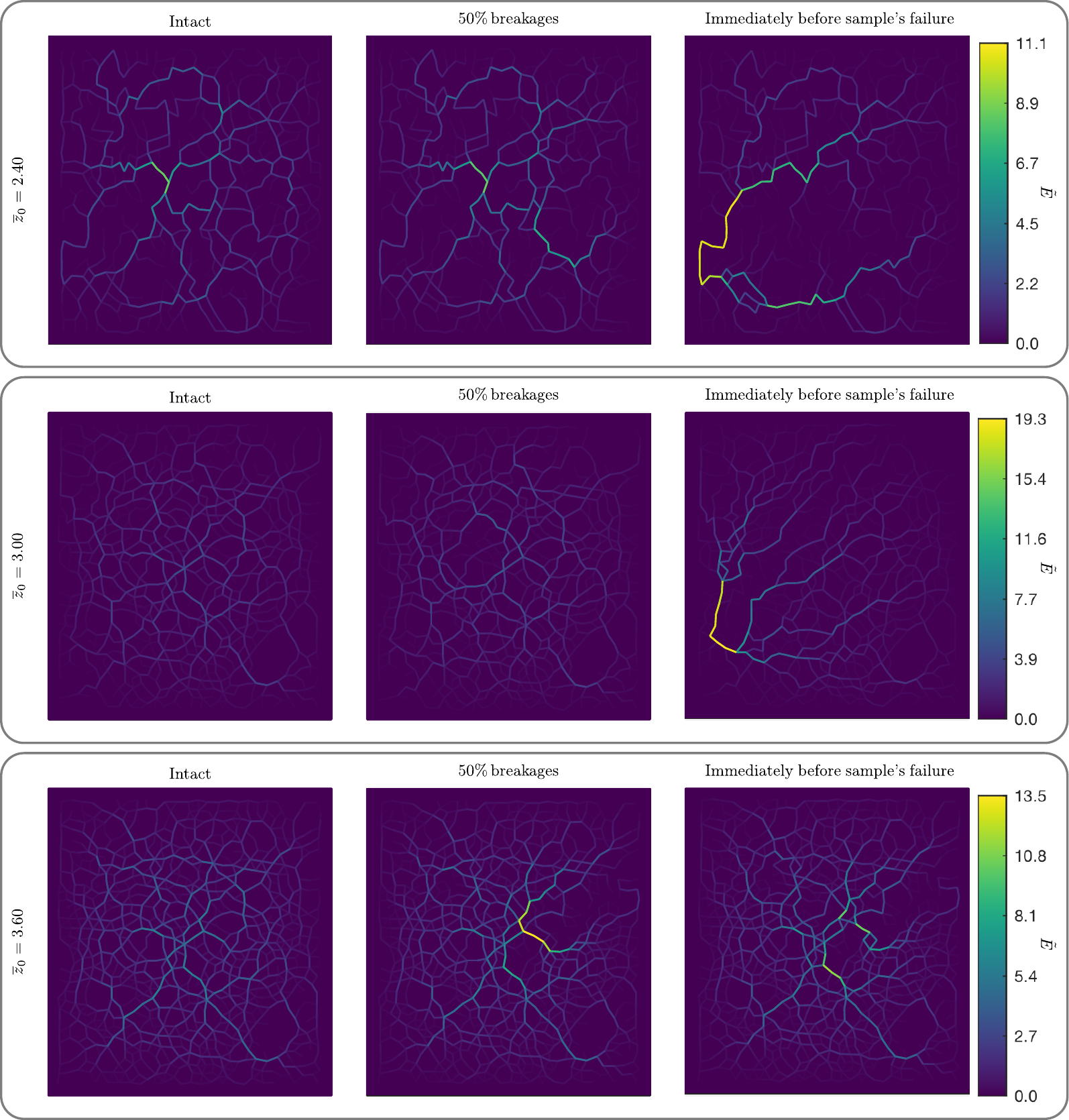} 
\caption{ Example images of the spatial distribution of normalized geodesic edge betweenness centralities $\tilde{E}$ (given by the color bar), which we plot at a particular applied strain $\varepsilon$ (see ``Mechanical testing protocol'' in Materials and Methods for a discussion of strain steps) for samples that are subject to compression.
The rows show samples with
(top) $\z_0 = 2.40$; 
(middle) $\z_0 = 3.00$; 
(bottom) $\z_0 = 3.60$; 
Within each row, we show the progression (of strain steps) in $\varepsilon$ from
(left) initial, intact networks $G_0$ with adjacency matrix ${\bf A}_0$; to
(center) the step at which $50\%$ of breakages have occurred 
(with $\varepsilon \approx 3.39 \%$, $\varepsilon \approx 1.68 \%$, and $\varepsilon \approx 1.90\%$ from top to bottom); and finally to
(right) the strain step immediately before a system-spanning failure
(with $\varepsilon \approx 9.56 \%$, $\varepsilon \approx 3.66 \%$, and $\varepsilon \approx 1.95\%$ from top to bottom).
} 
\label{fig:Maps}
\end{figure*}

Damage occurs progressively through a sequence of tensile or compressive loading. In Fig.~\ref{fig:Maps}, we show examples of damage progression for three values (one per row) of $\z_0$. Within each row, a sample progresses from its initial, intact network $G_0$ (an unweighted and undirected graph) through an altered network at which approximately 50\% of its beams have failed, and then to the network immediately before the final failure. In the image immediate after the last one that we show in each row, there is no longer a set of beams that connects the top and bottom boundaries of the sample. We color each edge in a network according to the value of $\Ee_s$ at that strain step. 

Geodesic edge betweenness is spatially heterogeneous across a network, and we observe that large values (bright colors) can occur throughout a network. These locations shift both in space and in time, due to the disordered structure (which arises from geometry) of the lattice. 
By contrast, for a regular lattice, the importance of edges decreases with their distance from the geometric center of a system \cite{barthelemy_morphogenesis_2018}. The introduction of disorder --- such as by rewiring, addition, or removal of edges --- results in more complicated distributions and can lead to geographically central edges with smaller importance than elements that are farther from the geometric center \cite{lion_central_2017}.
Importantly, although the topologies of the networks underlying our lattices
are inherited from uniaxially compressed granular packings, we not observed a preferred orientation for edges with GEBC values that are above the mean.
Granular packings encode their preparation history in the form of anisotropic stresses \cite{majmudar_contact_2005,bililign2019}, but this anisotropy is not readily identifiable from the contact network (which is unweighted).

The GEBC values at a given strain step illustrate the broad distribution of values, as we observed in the exponential probability density function of $\Ee_0$ (see Fig.~\ref{fig:EBC}). 
Even in these small systems, some edges have values up to $20$ times the mean of the system; these are ones that are particularly important for connecting different parts of a network. Many other edges occur only infrequently as shortest-path connectors. The variations in spatial distribution along the rows of Fig.~\ref{fig:Maps} highlight the importance of the removed edges, as we emphasized in Fig.~\ref{fig:EBC}b. 
Importantly, although $\Ee$ tends to decrease with distance from the geometric center, this need not be true for specific samples. For the near-final networks (in the right-most column in Fig.~\ref{fig:Maps}) at $\z_0=2.40$ and $\z_0=3.00$, the maximum of $\Ee$ is located near the left boundary of the sample, rather than near the middle. In both cases, the largest values of $\Ee$ occur on edges that connect the top and bottom parts of the network, and these are also the next beams that will break (and lead to the final cascade of failures). 


\subsection{Geodesic edge betweenness centralities of failed edges}

\begin{figure*}
\center
 \includegraphics[width=\linewidth]{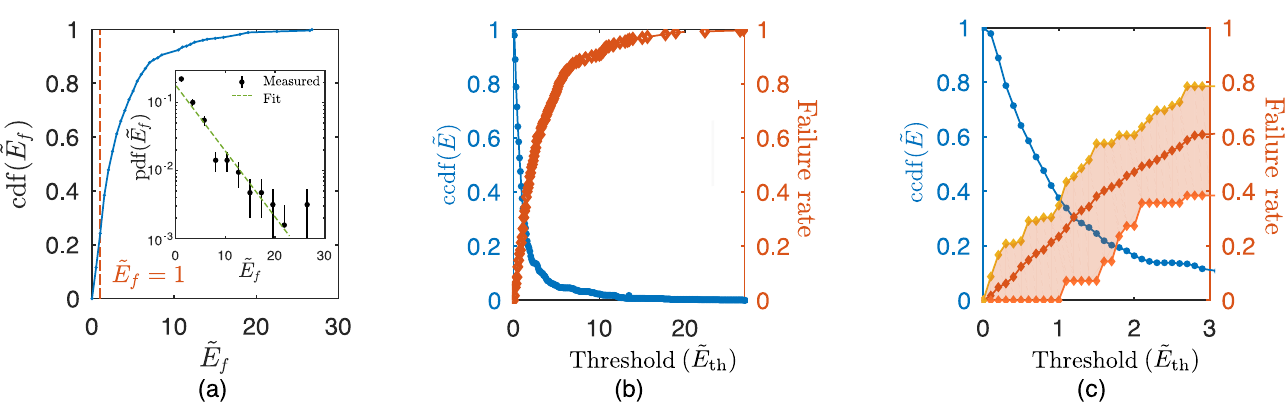} 
\caption{
(a) Cumulative distribution function of geodesic edge betweenness centrality of failed edges of all experiments. We show the probability density function in the inset. 
(b) Fraction of edges in the network for which $\Ee>\Ee_{\mathrm{th}}$ (blue dots, left axis) and fraction of failed beams for which $\Ee\leq\Ee_{\mathrm{th}}$ (orange diamonds, right axis). (c) Magnification of the crossover point between the CCDF and the failure rate (and the envelope of results of individual samples). 
} 
\label{fig:CDFs}
\end{figure*}

Such observations suggest that there is a correlation between large values of $\Ee$ and future failure locations. To assess the generality of this finding, for each breaking beam, we calculate the GEBC $\Ee_f$ during step $s-1$ immediately before its failure at step $s$. 
For all of our samples and for all non-large failure events (which we take to mean that no more than three beams are involved), we enumerate the immediately-preceding values of $\Ee$ for the failed edges. In Fig.~\ref{fig:CDFs}a, we show the cumulative distribution function (CDF) of this set of values, together with the corresponding probability density function (PDF) in the inset. We fit the PDF with an exponential with mean $\Ee_f^\ast \approx 10.3$ (with $R^2 \approx 0.96$).
There is a corresponding gradual increase for $\Ee_f \gtrapprox 10$ of the CDF, suggesting that few failing edges have a value that is significantly larger than the mean. We observe such large values of $\Ee$ only when the samples are near full failure; at this point, only a few paths are available to connect the top and bottom boundaries of the network. One can see this situation in the right column of Fig.~\ref{fig:Maps}. Focusing on $\Ee_f = 1$, we see that about $76\%$ of the breakages occur on edges with values of $\Ee_f$ that are above the mean. Because only a small subset of the network's edges have $\Ee>1$ (see the distribution in Fig.~\ref{fig:EBC}a), even the value of $\Ee$ alone is a valuable diagnostic for forecasting failure locations. 

We can refine this diagnostic by directly considering the population of edges that exceed a threshold value $\Ee_{\mathrm{th}}$. We illustrate this population by plotting the complementary cumulative distribution function (CCDF) on the left vertical axis in Fig.~\ref{fig:CDFs}b. Because the proportion of edges that satisfy $\Ee>\Ee_{\mathrm{th}}$ evolves after each edge removal and differs across initial networks, we choose each point of the curve to be the maximum value that we encounter among all networks. The success rate of this diagnostic is the fraction of failed beams that satisfy $\Ee > \Ee_{\mathrm{th}}$, and the failure rate is the fraction for which $\Ee \leq \Ee_{\mathrm{th}}$. We show the latter in orange diamonds in the right vertical axis of Fig.~\ref{fig:CDFs}b for all non-large failure events among all tested samples, regardless of the tensile or compressive nature of the applied loading. In Fig.~\ref{fig:CDFs}c, we focus on the point at which the CCDF and the failure-rate curves cross; this intersection occurs at $\Ee_{\mathrm{th}} \approx 1.1$, corresponding to a value on the CCDF curve (i.e., the fraction of edges for which $\Ee \gtrapprox 1.1$) of about $0.34$ and a failure rate of about $0.26$. This intersection point indicates that considering all edges with above-mean geodesic edge betweenness values provides a reasonable population of edges to consider, but one can choose other values in a trade-off between forecast failure rate and the fraction of examined edges.

The above general results exhibit sample-to-sample variation. To highlight this, we include an envelope of the failure rate in Fig.~\ref{fig:CDFs}c. To obtain this envelope, we determine a failure rate curve for each of the $14$ samples (see Materials and Methods). We obtain each curve by examining the failure events that occur on each initial intact network. For each threshold value, we track the best (lower point) and worst (upper point) failure-rate value among the $14$ curves. The envelope is the set of points between these lower and upper bounds for each threshold value. Although the scatter is non-negligible, for a threshold of $\Ee_{\mathrm{th}} = 1$, we still obtain success rates above 65\% for all samples.


\subsection{Test sensitivity and specificity}

Performing sensitivity and specificity analysis \cite{simundic_measures_2009} allows a more detailed determination of the suitability of using $\Ee > \Ee_{\mathrm{th}}$ to identify beams that are likely to fail. We define the outcome of this test as a true positive (TP), false positive (FP), true negative (TN), or false negative (FN) according to the state of the beam (failing or remaining intact). See our summary in Table~\ref{tab:TestDef}.

{\it Sensitivity} is defined as the probability of obtaining a positive test result for the population of failed beams (i.e., the proportion of true positives), so $\mathrm{sensitivity} = \mathrm{TP}/(\mathrm{TP} + \mathrm{FN})$.
Similarly, {\it specificity} is the probability of obtaining a negative test result for the population of intact beams (i.e., proportion of true negatives), so $\mathrm{specificity} = \mathrm{TN}/(\mathrm{TN} + \mathrm{FP})$.
These two measures quantify the success of our test for correctly identifying beams that will fail or remain intact.

\begin{figure}
\center
\includegraphics[width=\linewidth]{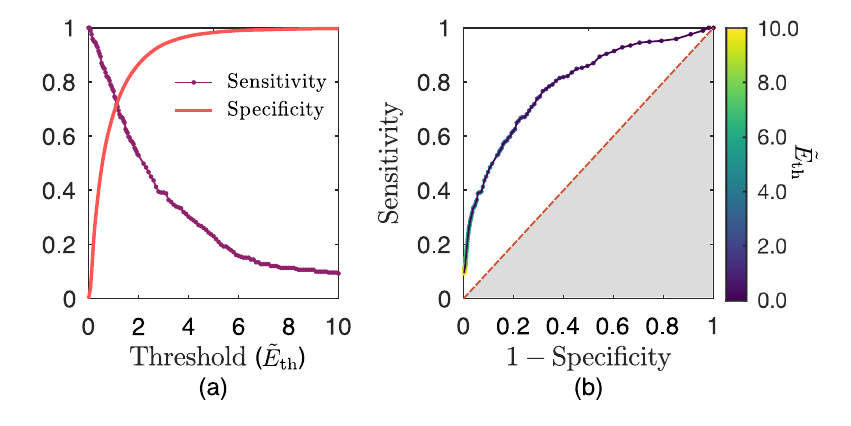} 
\caption{Evaluation of our test's accuracy. (a) Sensitivity and specificity versus the threshold $\Ee_{\mathrm{th}}$. (b) Receiver operating characteristic (ROC) curve summarizing the $(\mathrm{sensitivity}, 1-\mathrm{specificity})$ pairs that we obtain for different values of $E_c$. The dashed diagonal line indicates the behavior of a test that cannot discriminate between failing and intact beams.
} 
\label{fig:ROC}
\end{figure}

We calculate sensitivity and specificity in considering all non-large failure events of all experiments as a function of the threshold $\Ee_{\mathrm{th}}$, and we show the results in Fig.~\ref{fig:ROC}a. As expected, sensitivity and specificity show opposite trends: as one lowers the threshold, the true-positive fraction (sensitivity) increases, but so does the false-positive fraction, so that the specificity (i.e., the true-negative fraction) decreases. As one increases the threshold, the opposite occurs: we obtain a lower true-positive fraction (sensitivity decreases), and the false-positive fraction decreases (specificity increases). There is a crossover between sensitivity and specificity at $\Ee_{\mathrm{th}} \approx 1.1$, which is close to the value $1$ that we used above. 

Computing a receiver operating characteristic (ROC) curve \cite{zweig_receiver-operating_1993} provides additional insight into the choice of $\Ee_{\mathrm{th}}$. As we show in Fig.~\ref{fig:ROC}b, we measure sensitivity and specificity as a function of $\Ee_{\mathrm{th}}$. A test with perfect forecasting of failing versus intact beams would go through the upper-left corner (in which sensitivity and specificity are both $1$), and a test without any predictive power would follow the dashed diagonal line. (Anything below this line gives a result that is worse than random guessing and indicates a test direction that is opposite to what should have been chosen.)

To obtain a global estimate of the accuracy of the test that goes beyond visual examination, we compute the area under the curve of the ROC curve. This ranges from $0.5$ (no discrimination) to $1$ (perfect accuracy). The value for the curve in Fig.~\ref{fig:ROC}b is $0.79$, indicating a good capability of our test to discriminate between beams that will remain intact versus those that will fail.

\begin{table}
\centering
\caption{Definition of the outcome of a test.}
\begin{tabular}{c| c c}
Test: Is $\Ee>\Ee_{\mathrm{th}}$?     &  Beam fails  & Beam does not fail\\ 
\hline
Positive &  True Positive (TP)   & False Positive (FP) \\ 
Negative &  False Negative (FN)   &  True Negative (TN)\\ 
\hline
\end{tabular}
\label{tab:TestDef}
\end{table}


\section{Discussion} \label{discussion}

\subsection{Other network diagnostics and approaches for forecasting failures}

Motivated by the results in \cite{smart_effects_2007} and the geometric origin of our samples, we have focused on using geodesic edge betweenness centrality for forecasting failures in them. However, other network measures are also worth considering as possible diagnostics for forecasting failure locations. In particular, it is desirable to take advantage of the fact that the various flavors of betweenness are correlated with other quantities in certain types of networks. In some networks, for example, geodesic node betweenness can scale approximately with node degree \cite{barthlemy_betweenness_2004}. To give another example, Scellato \textit{et al.} \cite{scellato_backbone_2006} studied the relation between GEBC and a quantity known as {\itshape information centrality} in networks based on the road systems of several cities. 
For the edge $e_{ij}$ of a network, information centrality is 
\begin{equation}
	J_{ij} = \frac{F[G]-F[G']}{F[G]}\,,
\end{equation}
where 
\begin{equation}
	F[G] = \dfrac{1}{N(N-1)}\sum_{i,j=1,\ldots,N;\,i\neq j}\dfrac{d_{ij}^{\mathrm{Eucl}}}{d_{ij}}
\end{equation}
is the 
{\itshape efficiency} of an $N$-node graph $G$, the graph $G'$ results from removing edge $e_{ij}$ from $G$, the quantity $d_{ij}$ is the distance between nodes $i$ and $j$ (e.g., from the shortest number of steps between $i$ and $j$ in an unweighted graph), and $d_{ij}^{\mathrm{Eucl}}$ is the Euclidean distance between those nodes.

We investigate the relation between geodesic edge betweenness centrality and information centrality using a modified expression for efficiency, where we set $d_{ij}^{\mathrm{Eucl}}=1$ for all pairs $(i,j)$ with $i \neq j$. To calculate information centrality, we use a code from \cite{noauthor_brain_nodate}. Information centrality gives an indication of the perturbation of transmission across a network when one removes an edge. In other words, we ask the following question: how harmful is a beam failure for connections across a sample? When considering all edges of all of our networks (both initial and altered), we obtain a correlation (with a Spearman rank-correlation coefficient of $0.55 \pm 0.014$) of information centrality with geodesic edge betweenness centrality. For failed edges, the Spearman correlation coefficient ($0.77 \pm 0.01$) is even larger. Motivated by these calculations, we checked and confirmed that considering values above the mean for information centrality yields similar results as using geodesic edge betweenness centrality as a test for potential failures. Therefore, information centrality is an alternative to geodesic edge betweenness centrality to probe systems for likely failure locations. Measures based on shortest paths are not always highly correlated with each other (and the extent of such correlation also depends on network type) \cite{lee2014}, so different measures related to betweenness can give complimentary insights.

\subsection{Comparison with other diagnostics}
To disentangle the roles of physical and geometric effects, we evaluate the importance of introducing physical considerations by repeating our analysis using an RFN-based test. (See ``Random-fuse network (RFN) model'' in Materials and Methods.)
In place of {$\Ee$}, we determine the current $\tilde{I}$ that flows through a network.
 As with GEBC, we determine the fraction of edges in a network with a 
 current above a threshold current $\tilde{I}_{\mathrm{th}}$ and the fraction of failing beams for which $\tilde{I}_f\leq \tilde{I}_{\mathrm{th}}$ (where $\tilde{I}_f$ denotes the current of an edge during step $s-1$ immediately before its failure at step $s$).

For threshold values in the range $[0,2.5]$, we find that this test performs somewhat better than the test using GEBC (see Fig.~S1a). 
However, in the range $[0,1.4]$, the fraction of edges with a current above the threshold is slightly larger than the fraction of edges with a GEBC above the same threshold. Consequently, the trade-off between the forecast failure rate and the fraction of possible edges differs between the two tests for the same threshold value. This trade-off diminishes the advantage.

We also find that the current and GEBC values of the failed edges are positively correlated, with a Pearson correlation coefficient of $R \approx 0.81$ (see Fig.~S1b). 
This suggests that failing beams, most of which have a larger stress than the mean value in the network (i.e., $\tilde{I}_f>1$ for these edges) lie on many shortest paths, explaining the small improvement in forecast capability from using the RFN-based test. 
This finding is similar to the results of Kollmer {\it et al.} \cite{kollmer2019betweenness}, who observed in 2D packings of frictional particles that ones with large {\it node} betweenness centralities are statistically likely to be highly stressed. Consequently, it is appropriate to examine betweenness centralities (of both nodes and edges) to capture important mechanical properties of physical systems.
 
Interestingly, as we show in Fig.~S2, the two tests do not systematically misdiagnose failure locations (i.e., yield false-negative outputs) for the same edges. Using the RFN-based test, we observe a correlation between beam angle and current flow (see Fig.~S3a) and find a bias towards misdiagnosed edges that are roughly perpendicular to the loading direction (see Fig.~S3b). It appears that failures occur at edges at all distances from the boundaries (see Fig.~S4a). However, for some threshold values, most GEBC-misdiagnosed edges tend to occur near boundaries (see Fig.~S4b), where large values of GEBC are less frequent.

The dependency of geodesic edge betweenness centrality with distance from a sample's geometric center, even altered by the presence of disorder, is an important feature of networks that are embedded in a plane. To test whether we can circumvent this limitation of the test, we calculate geodesic edge betweenness centralities on a collection of modified networks. For a given network $G$ and for each edge $e_{ij}$, we generate a duplicated network such that the edge $e_{ij}$ is at the geometric center. To construct such a graph, we first duplicate the network with mirror symmetries with respect to each boundary. We then calculate the geodesic edge betweenness centrality of $e_{ij}$ by considering only a section of this duplicated network that is approximately centered on $e_{ij}$.  We repeat this procedure for each edge of the network $G$ to obtain centrality values for the network, centered at that edge. Using this approach yields a cumulative distribution for $\Ee_f$ and a test ($\Ee_f>\Ee_{\mathrm{th}}$) success rate similar to the original networks. 
Indeed, while boundary edges can have large $\Ee$ values, 
the distributions of $\Ee$ are more homogeneous than in the original networks, such that the test can misdiagnose edges in other locations.
Consequently, the use of a duplicated network does not improve the forecasting ability of our approach.
 Developing methods to appropriately consider the role of boundaries remains a central question for planar graphs --- not only for granular materials, but also for other applications, such as determining high-traffic edges in road networks \cite{lee2014} and nutrient-transportation networks \cite{lee2017} --- and more generally in spatially-embedded networks.

In our comparison of these GEBC-based and RFN-based tests, we observe a correlation between the physical and geometric properties of failing beams. Interestingly, a  test that includes a minimal set of physical ingredients (i.e., the RFN-based test) performs only somewhat better than a test (the GEBC-based test) that is based on geometric considerations. Because these two tests have rather different limitations --- with less successful forecasting of near-perpendicular edges versus near-boundary edges --- it is useful to employ both as complementary approaches. Finally, it is worth noting that although the RFN-based test is faster computationally than the GEBC-based one, neither approach requires significant computational resources for the system sizes that we consider in this study.

\section{Conclusions} \label{conclusions}

The idea, proposed in papers such as \cite{smart_granular_2008} and reviewed in \cite{papadopoulos_network_2018} in the context of granular and particulate systems, to use network analysis to achieve insights on novel physical systems seems very promising for studies of fracture. 
In this paper, we explored the application of centrality measures (based on shortest paths) to forecast failure locations in physical samples. Many other tools from network analysis, such as those based on exploration of mesoscale structures, also promise to yield fascinating insights into investigations of physical networks. In particular, examining how they can contribute to forecasting not only where, but also when, failures occur in disordered networks is a central point for future studies.

In the present investigation, we found that calculations based on shortest paths can be very helpful for forecasting failure locations in disordered lattices. Specifically, calculating the geodesic betweennesses of the edges in a network permits one to assess which edges are more prone to failure than others. Considering only edges with values above the mean geodesic edge betweenness of a network allows one to discard a large fraction of edges as unlikely failure locations. This feature of the test makes it very valuable, particularly as it avoids a detailed analysis of energetics. 

Combined with \cite{smart_effects_2007,kollmer2019betweenness}, our work provides evidence that betweenness centrality successfully identifies physical properties in both granular packings and lattices that are derived from them. Similarly, analyses inspired by rigidity percolation in granular materials have identified that our disordered lattices undergo a ductile--brittle failure transition as a function of connectivity $\z_0$, as determined by counting degrees of freedom and constraints \cite{berthier_rigidity_2018}. 
However, we have focused on testing the sensitivity and specificity of our approach in the context of disordered lattices that we generated from force networks in quasi-2D granular packings, and this does not ensure its success for other network structures.  Indeed, it has been established that the origin of a disordered structure --- whether from numerical spring networks, frictionless jammed sphere packings, or diluted networks --- can strongly affect elastic responses and the rigidity transition \cite{ellenbroek_non-affine_2009,ellenbroek_rigidity_2015}. Therefore, we expect that both the granular origins of our samples (and hence their geometry) may affect the particular failure behavior that we have observed in this study. Collectively, these investigations (and ours) point towards a need to understand the importance of network topology and geometry themselves, regardless of their manifestation as a granular packing or a lattice. This is an important step towards distinguishing between geometric effects and system-specific physical effects. Therefore, an important future direction is to examine networks obtained by other means, such as an over-constrained granular packing, a biological system such as leaf-venation patterns, or by randomly pruning a crystalline lattice.

Our work also opens the door for structure design and the purposeful setting of desired failure locations. One can build particular network topologies into designed materials that permit the constraining of failures to regions of a sample or, by contrast, promote desirable patterns of damage spreading to ensure the robustness of structures.


\section{Methods} \label{sec:methods}

\subsection{Experimental samples} \label{sec:samples}

We conduct experiments on a set of disordered structures that we derived from experimentally-determined force networks in granular materials, as done in Berthier \textit{et al.} \cite{berthier_rigidity_2018}. The methodology to create these experimental samples is inspired by the work of \cite{driscoll_role_2016,goodrich_principle_2015,reid_auxetic_2018, ellenbroek_non-affine_2009,ellenbroek_rigidity_2015}, who performed a similar process numerically. We begin from observed force-chain structures in a quasi-2D photoelastic granular material. The granular packings consist of $N=824$ bidisperse circular discs (of two distinct radii $r_1$ and $r_2$, with $r_1/r_2 = 1.4$) in approximately equal numbers, as shown in Fig.~\ref{fig:method}a. We uniaxially load each packing under a series of finite displacements of one wall, generating multiple realizations of both packings and force networks. 

Using an open-source photoelastic solver \cite{daniels_photoelastic_2017,jekollmer_pegs:_2018}, we identify all load-bearing contacts in a system, yielding a network of physical connections between particles that we use to generate a disordered lattice (see Fig.~\ref{fig:method}b). We construct a network by assigning each particle center as a node of a graph $G$ and then placing an edge between two nodes wherever we observe a load-bearing contact. The network is associated with an $N \times N$ binary (i.e., unweighted) adjacency matrix ${\bf A}$, with elements
\begin{equation}\label{matrix}
A_{ij}=\left\{
     \begin{array}{ll}
     1\,, \, \text{if particles}\,\, i\,\,\text{and}\,\, j \,\, \text{are load-bearing}\,,\\
     0\,,\, \text{otherwise}\,.\\
     \end{array}
     \right.
 \end{equation} 
We showed an example network in Fig.~\ref{fig:method}c. It is undirected because each contact is bidirectional, and its associated adjacency matrix is therefore symmetric about the diagonal ($A_{ij} = A_{ji}$).

We laser-cut the physical samples from acrylic plastic sheets (with an elastic modulus of about $3$ GPa) of thickness $h = 3.17$~mm. Each edge becomes a beam of width $1.5$~mm; beams intersect at crosslinks that correspond to the centers of particles (i.e., the nodes), and the particles' radii sets the length of the beams. We adopt the term ``crosslink'' from the study of fiber networks \cite{zhang_fiber_2017,broedersz_modeling_2014}, which
consist of  filaments (bonds) that are bound via crosslinkers that either allow energy-free rotations or associate angular variations to a finite cost of energy (as ``welded'' crosslinks). We showed an example sample in Fig.~\ref{fig:method}d; note that samples that we construct multiple times based on the same mathematical networks by cutting from different sheets of material are not perfectly identical due to small details of processing during cutting. 

A simple characteristic of a network is its {\itshape mean degree} $\z$ (also known as the ``connectivity'' or ``coordination number''), which is equal to the mean number of edges per node. That is,
\begin{equation}
	\z = \frac{1}{2N}\sum_{i,j}^N A_{ij}\,.
\end{equation}
It is known that the bulk properties of amorphous solids \cite{wyart2005, berthier_rigidity_2018} are influenced strongly by $\z$.
We use the subscript $0$ denote initial networks (i.e., networks before any subsequent modifications from lattice beam failures).
We study $6$ different initial networks, with mean degrees $\z_0 = \{2.40, 2.55, 2.60, 3.00, 3.35, 3.60\} \pm 0.02$, which we draw from two different initial granular configurations. 
 We do a total of $14$ experiments, which we test each network at least once in compression and once in tension; for the networks with $\z_0 = 2.60$ and $\z_0 = 3.00$, we do an additional tensile test on 
a second set of fully intact samples. 
To obtain a sample that is as close as possible to the isostatic value $\z_{\mathrm{iso}} =  3.00$ of an infinite friction packing \cite{henkes_rigid_2016}, we prune a network that initially has a value of $\z_0 = 3.60$ by progressively removing its contacts with the smallest force values. 

\subsection{Mechanical testing protocol} \label{sec:exper}

We perform compression and tension tests using an Instron 5940 Single Column system with a $2$~kN load cell. We use a displacement rate of  $1.0$~mm/min for tension experiments and $1.5$~mm/min for compression experiments. In compression, we confine a sample between two parallel acrylic plates to constrain out-of-plane buckling. We record each experiment using a Nikon D850 digital camera at a frame rate of 24 or 60 fps. During the course of each experiment, beams break throughout a sample as damage progresses. Using the time series of measured compressive and tensile forces, we identify each {\it failure event}, which corresponds to a set of one or more breakages that occur simultaneously. Our frame rates are insufficient to distinguish multiple, successive breakage events that occur within a single failure event, but they are sufficient to easily separate the failure events from each other. In all cases, we are able to determine the locations of individual beam failures by examining the images collected immediately following a recorded drop in force. The failure events occur sequentially, deteriorating the structure until complete failure of a sample. This corresponds to having a crack going through the sample from one lateral side to the other, such that there is no set of beams that connects the top and bottom boundaries.

As damage progresses, the adjacency matrix ${\bf A}$ (and the associated network $G$; see \eqref{matrix}) that encodes the network structure changes following each failure event. When the beam that connects nodes $i$ and $j$ fails, we set $A_{ij} = A_{ji} = 0$ to record this event. We thus do a series of computations on networks that are based on measurements at a particular strain step $s$, which is associated to an applied strain value $\varepsilon$. We distinguish between the initial network $G_0$ (with adjacency matrix ${\bf A}_0$), which is associated with the fully intact sample, and altered networks $G_s$ (with associated adjacency matrices ${\bf A}_s$).

Note that, as characterized in \cite{berthier_rigidity_2018}, both tensile and compressive loading of samples with $\z_0 < \z_{\mathrm{iso}}$ will fail from breakages that are well-separated in time and are spatially spread in a sample (i.e., ductile-like failure). By contrast, it was shown in \cite{berthier_rigidity_2018} that for $\z_0  > \z_{\mathrm{iso}}$, a few temporally separated breakages take place before the samples break
abruptly and all of the failed beams are localized, forming a narrow crack (i.e., brittle-like failure).
Therefore, for the samples with $\z_0 = 3.35$ and $\z_0 = 3.60$, the deterioration of a sample's structure occurs via both small (one to three breakages at a time) and large (more than three simultaneous breakages) failure events. In our analysis, we remove corresponding edges from the networks as failures take place, and we then perform fresh calculations of centrality.
Our analysis of failure locations excludes the large events, because we are specifically interested in the progression of failures. Our results are qualitatively similar for samples tested in tension versus in compression, so we do not distinguish between these two loading conditions in our analysis.


\subsection{Geodesic edge betweenness centrality (GEBC)} \label{sec:EBC}

Because failures in our samples consist mostly of breaking beams (rather than the thicker crosslinks), we focus on an edge-based counterpart of geodesic node betweenness centrality \cite{girvan_community_2002}. This measure gives insight into the importance of edges in a network in terms of how often they are on shortest paths between origin and destination nodes. Considering an edge $e_{ij}$ that links nodes $i$ and $j$ in a graph $G$, we calculate a symmetric {\itshape geodesic edge betweenness centrality}  matrix based on the fraction of shortest paths that traverse an edge when considering all origin--destination pairs of nodes in a network (including nodes $i$ and $j$) \cite{barthelemy_morphogenesis_2018}: 
\begin{equation}\label{ebc-def}
	E_{ij} = \sum_{s\neq t} \frac{\sigma_{st}(e_{ij})}{\sigma_{st}}\,,
\end{equation}
where $\sigma_{st}$ is the number of shortest paths from node $s$ to $t$, and $\sigma_{st}(e_{ij})$ is the number of those paths that include the edge $e_{ij}$. We compute $E_{ij}$ using open-source code from \cite{noauthor_brain_nodate} (which uses an algorithm that is a slight modification of the one in \cite{brandes_faster_2001}). This measure can be computationally costly for large networks. The computation time is ${\cal O}(Nm)$ for sparse networks, where $N$ and $m$ denote the numbers of nodes and edges, respectively, of a network \cite{girvan_community_2002}. All of our graphs $G$ are undirected and unweighted, but one can also study notions of edge betweenness centralities for directed and weighted graphs.

It is common to normalize $E_{ij}$ by $\tfrac{1}{2}N(N-1)-1$ (i.e., by the number of edges, other than the one under consideration) \cite{chen_preliminaries_2015} or by $(N-1)(N-2)/2$ (i.e., the number of node pairs) \cite{barthelemy_morphogenesis_2018} to ensure that geodesic edge betweenness values lie between $0$ and $1$. However, because we will compare the relative importance of edges to others in a given network and as successive edge removals occur, we use a different normalization. In our calculations, for a given network at strain step $s$ and characterized by its adjacency matrix ${\bf A}_s$ (where $s=0$ for the initial network), we define the normalized geodesic edge betweenness matrix $\tilde{{\bf E}}_{s} = {\bf E}_{s}/\E_{s}$, where $\E_{s}$ is the mean over all edges of the network $G_s$. To study the importance of the failing beams, for each strain step, we compute the matrix $\tilde{\bf E}_{s}$ and extract the values $\Ee_f$ of the edges that fail in the next failure event.


\subsection{Random-fuse network (RFN) model} \label{sec:RFN}

We create a RFN \cite{de_arcangelis_random_1985,kahng_electrical_1988} in which each fuse matches an edge of the network for one of our samples.
All fuses have identical conductance, because the beams have identical thickness.
We load the top and bottom boundaries by applying a fixed voltage to the top nodes and connecting the bottom nodes to ground (zero voltage). We determine edge voltages and their associated currents by solving Kirchhoff's laws. Analogous to our examination of GEBC, we normalize the current by the mean over all edge currents in a network. We define a current matrix at strain step $s$ by $\bold{\tilde{I}}_s = \bold{I}_s/\overline{I}_s$, where $s=0$ denotes the initial network and associated quantities. We then proceed as with GEBC: for each strain step $s$, we calculate the matrix $\tilde{\bf I}_s$ and extract the normalized current (i.e., stress) $\tilde{I}_f$ of the edges that fail in the next failure event.\\

\textbf{Acknowledgements:} We gratefully acknowledge Jonathan Kollmer for sharing the granular force network data that was collected in \cite{berthier_rigidity_2018}. This research was supported by the James S. McDonnell Foundation. We thank two anonymous referees for their many helpful comments, and we are also particularly thankful to Referee 2 for the suggestion to use the RFN-based test to perform a deeper analysis.

\renewcommand{\thefigure}{S\arabic{figure}}
\setcounter{figure}{0}

\section*{Supplementary Information}
\subsection*{Random-fuse network (RFN) current versus geodesic edge betweenness (GEBC) tests of failure locations}

\begin{figure}
\center
\includegraphics[width=\linewidth]{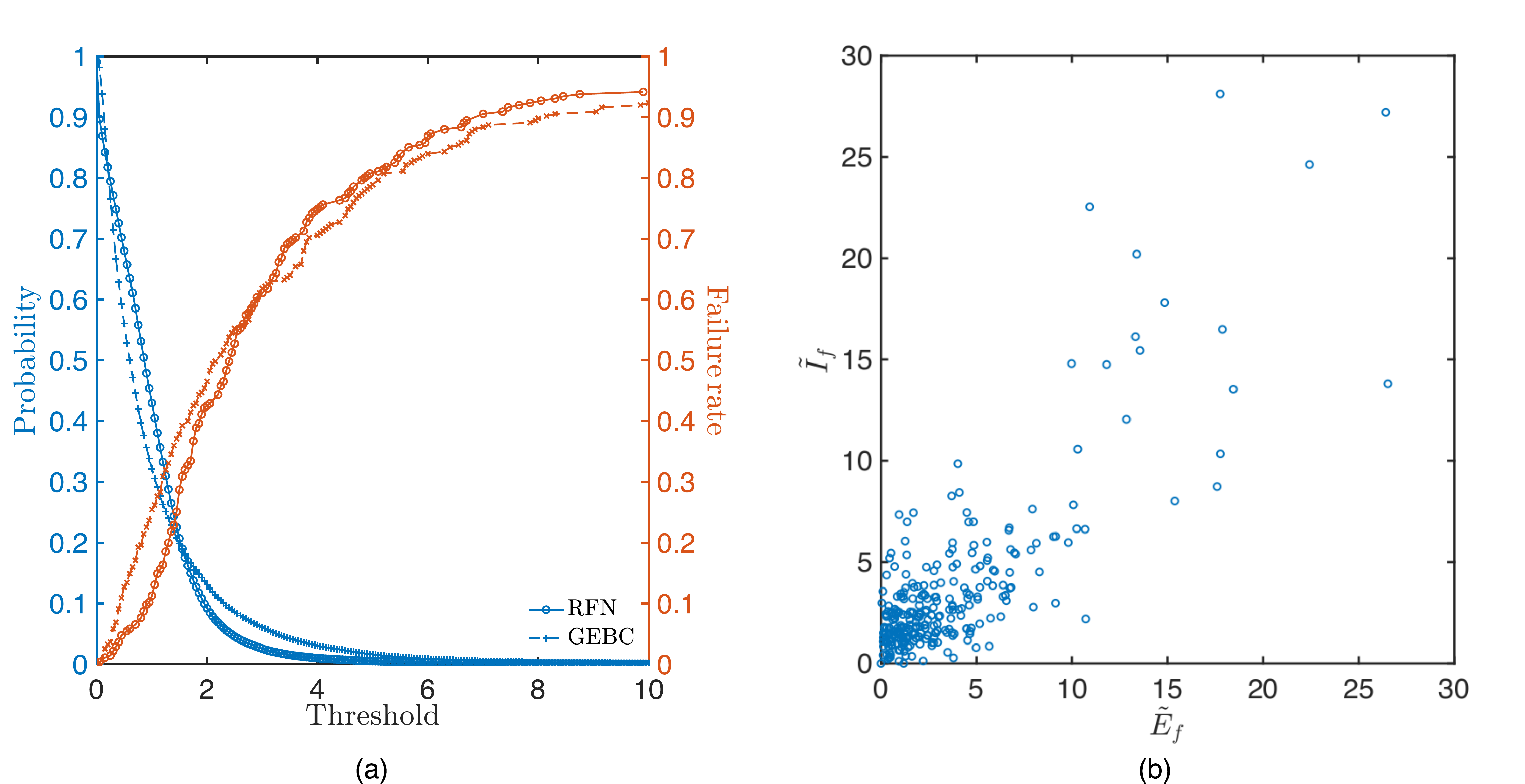}
\caption{(a) Comparison of the ability of the tests to forecast failure locations using currents from a random-fuse network (RFN; curves with circles) and geodesic edge betweenness centrality (GEBC; curves with crosses). On the left axis (in blue), we show the fraction of edges in a network for which the current is above a certain threshold (i.e., $\Tilde{I}>\Tilde{I}_{\mathrm{th}}$) and the fraction of edges for which GEBC is above a certain threshold (i.e., $\Tilde{E}>\Tilde{E}_{\mathrm{th}}$). On the right axis (in orange), we show the fraction of failed beams for which $\Tilde{I}\leq\Tilde{I}_{\mathrm{th}}$ and $\Tilde{E}\leq\Tilde{E}_{\mathrm{th}}$. The axis label ``Threshold'' indicates $\Tilde{I}_{\mathrm{th}}$ for the RFN-based test and $\Tilde{E}_{\mathrm{th}}$ for the GEBC-based test. (b) Scatter plot between the current ($\Tilde{I}_f$) and GEBC ($\Tilde{E}_f$) of failed edges. The Pearson correlation coefficient between $\Tilde{I}_f$ and $\Tilde{E}_f$ is $R \approx 0.81$.}\label{fig:S1}
\end{figure}

To elucidate the role of geometry in capturing failure locations, we compare the forecasting ability of a test that uses current that we compute from a random-fuse network (RFN) model to that of a test that uses geodesic edge betweenness centrality (GEBC). In the left axis of Fig.~\ref{fig:S1}a left, we show the fraction of edges in a network with a current $\Tilde{I}$ (respectively, GEBC $\Tilde{E}$) that is larger than a threshold current $\Tilde{I}_{\mathrm{th}}$ (respectively, threshold GEBC $\Tilde{E}_{\mathrm{th}}$). On the right axis, we show the failure rate as the fraction of misdiagnosed edges, which we define as failed edges with a current (respectively, GEBC) that is less than or equal to the threshold current (respectively, threshold GEBC). That is, $\Tilde{I}_f\leq\Tilde{I}_{\mathrm{th}}$ (respectively, $\Tilde{E}_f\leq\Tilde{E}_{\mathrm{th}}$) as a function of the threshold.

To explain the small improvement that we obtain when using a model with physical considerations (the RFN-based test), we show in Fig.~\ref{fig:S1}b the relation between the current ($\Tilde{I}_f$) of failed edges and the GEBC ($\Tilde{E}_f$) of these edges. The Pearson correlation coefficient between these two quantities is $R \approx 0.81$.

\begin{figure}
\center
\includegraphics[width=0.5\linewidth]{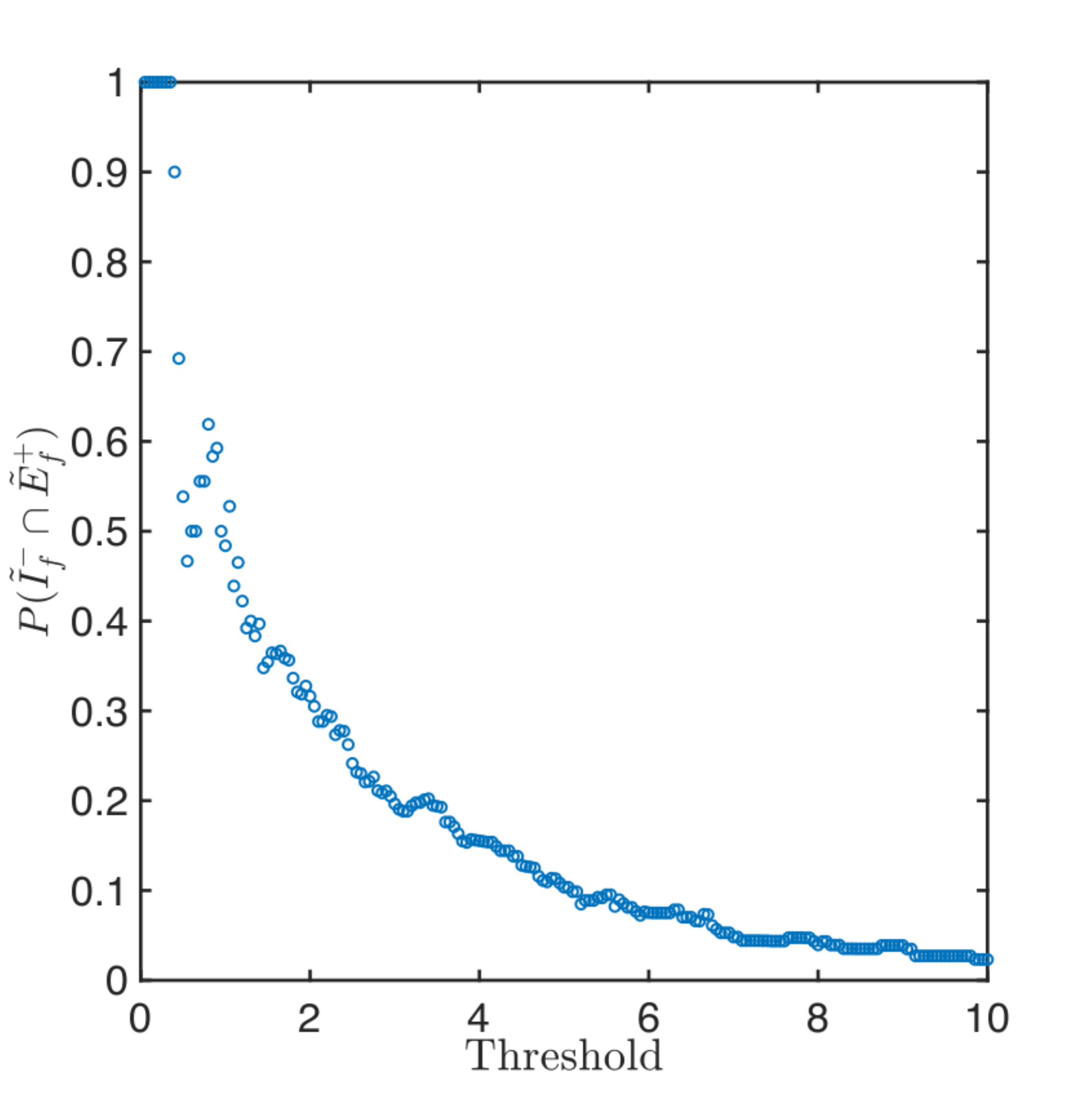}
\caption{Probability $P(\Tilde{I}_f^-\cap\Tilde{E}_f^+)$ that a misdiagnosed edge from the RFN-based test is diagnosed successfully using the GEBC-based test. The quantity $\Tilde{I}_f^-$ denotes an edge that is misdiagnosed using the RFN-based test (i.e., $\Tilde{I}_f\leq\Tilde{I}_{\mathrm{th}}$), the quantity $\Tilde{E}_f^+$ denotes an edge that is diagnosed successfully using the GEBC-based test (i.e., $\Tilde{E}_f>\Tilde{E}_{\mathrm{th}}$), and $\Tilde{I}_f^-\cap\Tilde{E}_f^+$ denotes an edge that is misdiagnosed using the RFN-based test but diagnosed correctly using the GEBC-based test. The axis label ``Threshold'' indicates the current $\Tilde{I}_{\mathrm{th}}$ for the RFN-based test and the GEBC $\Tilde{E}_{\mathrm{th}}$ for the GEBC-based test.  }\label{fig:S2}
\end{figure}

These two tests are not systematically misdiagnosing (obtaining a false negative outcome) failure locations for the same edges. We show this in Fig.~\ref{fig:S2}, where we plot the fraction of failed edges that are misdiagnosed using the RFN-based test and correctly diagnosed using the GEBC-based test as a function of the threshold. (Depending on the test, this is either a current threshold or a GEBC threshold.)

\subsection*{Dependency of current on orientation}

\begin{figure}
\center
\includegraphics[width=\linewidth]{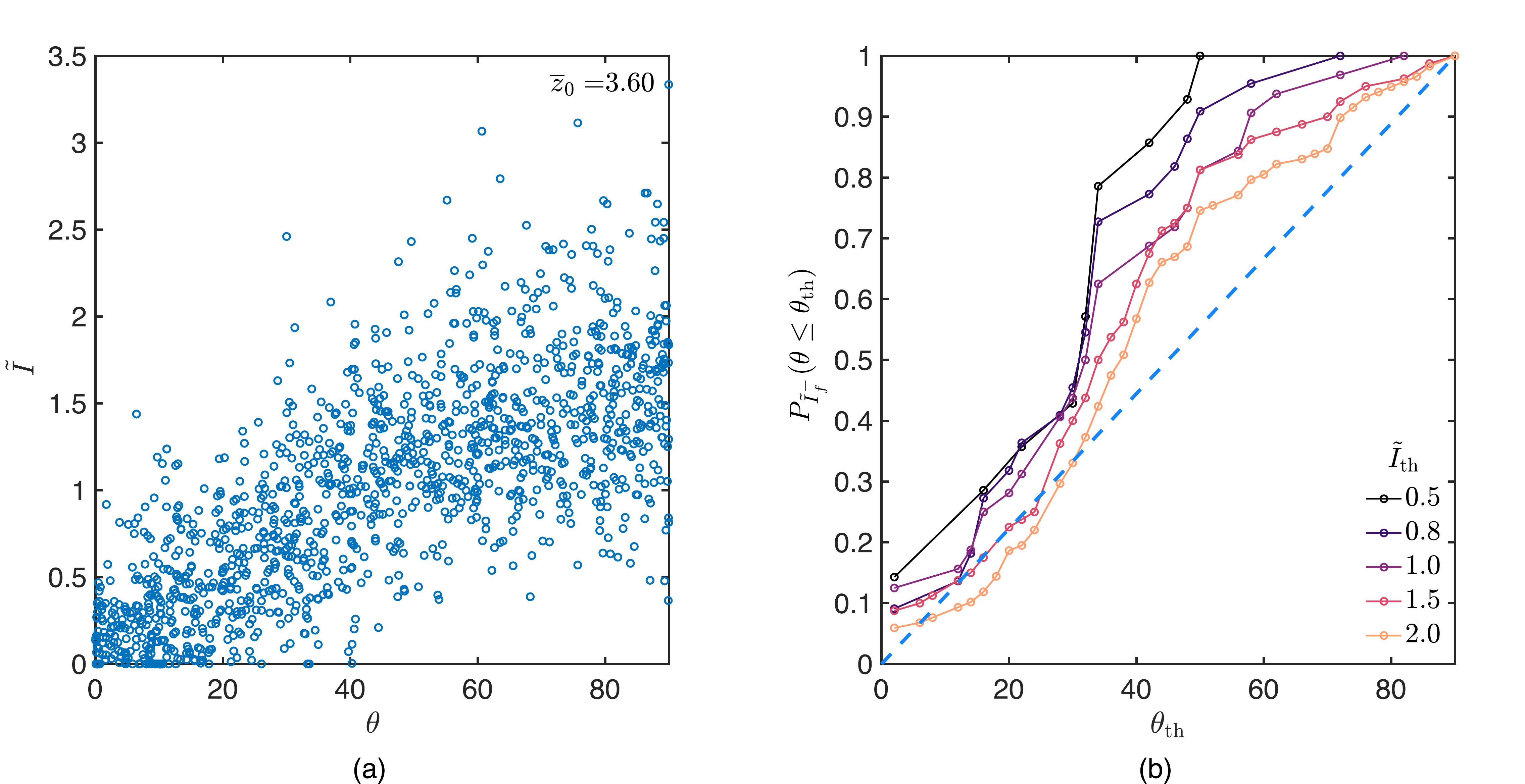}
\caption{(a) For the initial network with connectivity $\overline{z}_0 = 3.60$, we show a scatter plot between  edge current $\Tilde{I}$ and the edge orientation $\theta$ between $0$ degrees and $90$ degrees.
(b) For different values of the current threshold, we show the probability $P_{\Tilde{I}_f^-}$ that a misdiagnosed failed edge (i.e., with a current of $\Tilde{I}_f\leq\Tilde{I}_{\mathrm{th}}$), which is denoted by $\Tilde{I}_f^-$, has an orientation $\theta$ that is less than or equal to a threshold value $\theta_\mathrm{th}$. The dashed blue line indicates the theoretical behavior of misdiagnosed edges with no preferred orientation.
}\label{fig:S3}
\end{figure}

Consider a special case of a perfectly ordered squared lattice with identical resistors on each edge and a voltage difference that we apply across the top and bottom boundaries. In this configuration, no current flows in the horizontal edges. (This contrasts with the situation for GEBC, which is orientation-independent.) Therefore, we expect that --- to an 
 extent that depends on the amount of structural disorder --- any anisotropy remains imprinted in the current flow in the system. Indeed, as we see in Fig.~\ref{fig:S3}a, there is a strong dependence of the edge voltage on the edge angle (between $0$ and $90$ degrees) with respect to the vertical loading for a network with connectivity (i.e., mean degree) $\overline{z}_0 = 3.60$. The Spearman rank-correlation coefficient is $\rho \approx 0.77$.

 To determine the effect of this dependency on the forecasting ability of the RFN-based test, we show in Fig.~\ref{fig:S3}b, for various current thresholds $\Tilde{I}_{\mathrm{th}}$, the fraction $P$ of misdiagnosed edges ($\Tilde{I}_f^-$) that have an orientation with an angle $\theta$ that is smaller than a threshold $\theta_{\mathrm{th}}$. We thereby identify a bias towards misdiagnosing edges that are almost perpendicular (i.e., those with small $\theta$) to  the loading vertical direction.

\subsection*{Dependency of GEBC on edge distance to the boundaries}

Because one computes the GEBC of an edge based on the fraction of shortest paths that traverse an edge when considering all origin--destination pairs of nodes in a network, we expect edges near boundaries to have smaller values of $\Tilde{E}$, and we hence expect that they will be misdiagnosed more frequently by a GEBC-based test than edges that are located near the center of a network. To assess this possible limitation of our GEBC-based test, we calculate (in terms of the number of edges) the normalized shortest-path distance $\delta$ to reach a node on a boundary starting from a given node in the network. The normalized shortest-path distance ranges from $0$ (for a node on a boundary) to $1$ (if its shortest-path length to a boundary is the longest in a network). Therefore, nodes near boundaries have values of $\delta$ that are close to $0$, and those that are near the center of a network have values that are close to $1$. Specifically, the normalized shortest-path distance for a node $i$ in a network $G$ is 
\begin{equation} \label{dist}
    \delta_i = 1-\min_{j\in\mathcal{B}}\left( \frac{d^{\max} - d_{ij}}{d^{\max}} \right) \,,
\end{equation}
where $\mathcal{B}$ is the set of boundary nodes and $d^{\max}$ is the distance of a longest path between a node and a boundary in the network. We define the boundary $\mathcal{B}$ as the set of nodes that are located within one radius (where the radius is equal to half of the length of the longest beam in the network) from the nodes with highest or lowest vertical position (i.e., top or bottom nodes) and highest or lowest horizontal position (i.e., leftmost or rightmost nodes).

\begin{figure}
\center
\includegraphics[width=\linewidth]{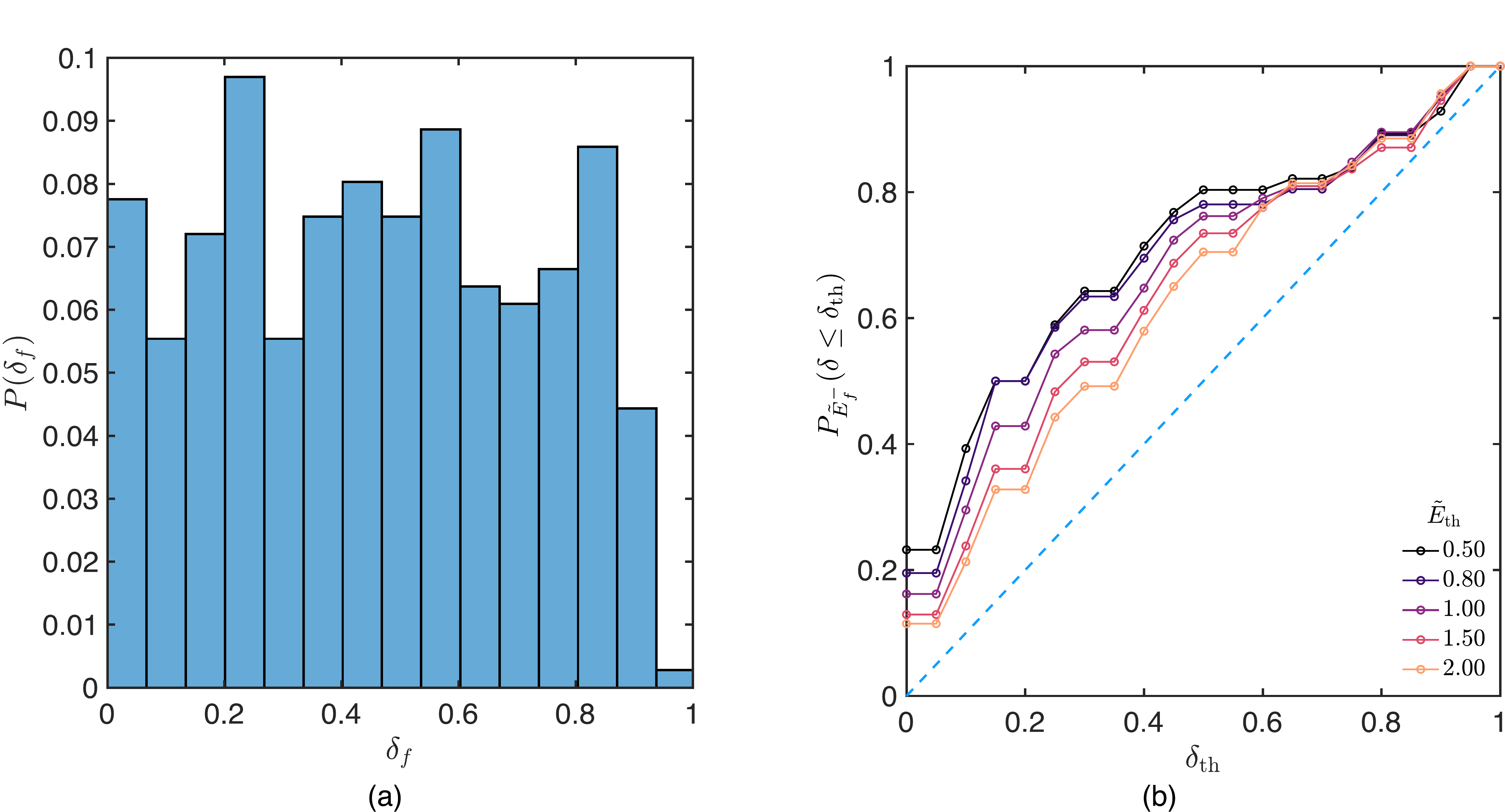}
\caption{(a) Histogram of $\delta_f$, the normalized shortest-path distance of failing edges to a boundary. (See the definition in the main text.) (b) For different GEBC threshold values $\Tilde{E}_{\mathrm{th}}$, we show the probability that a misdiagnosed failed edge (i.e., ones with with a GEBC of $\Tilde{E}_f\leq\Tilde{E}_{\mathrm{th}}$), which we denote by $\Tilde{E}_f^-$, is located at a distance $\delta_f$ that is less than or equal to a threshold value $\delta_\mathrm{th}$. The dashed blue line indicates the theoretical behavior of misdiagnosed edges with no preferred (normalized shortest-path) distance to the boundary in a network. }\label{fig:S4}
\end{figure}

For each failing edge, we compute the normalized shortest-path distance for both nodes that are associated with the edge, and we take the smaller value to define $\delta_f$ for each such edge. In Fig.~\ref{fig:S4}a, we show the histogram of distances $\delta_f$; and we thereby highlight that failing edges occur at all (normalized shortest-path) distances from the boundaries. However, as we show in Fig.~\ref{fig:S4}b, most misdiagnosed edges tend to occur near boundaries.

\subsection*{A comparison between the RFN model and GEBC on a small network}

\begin{figure}
\center
\includegraphics[width=0.5\linewidth]{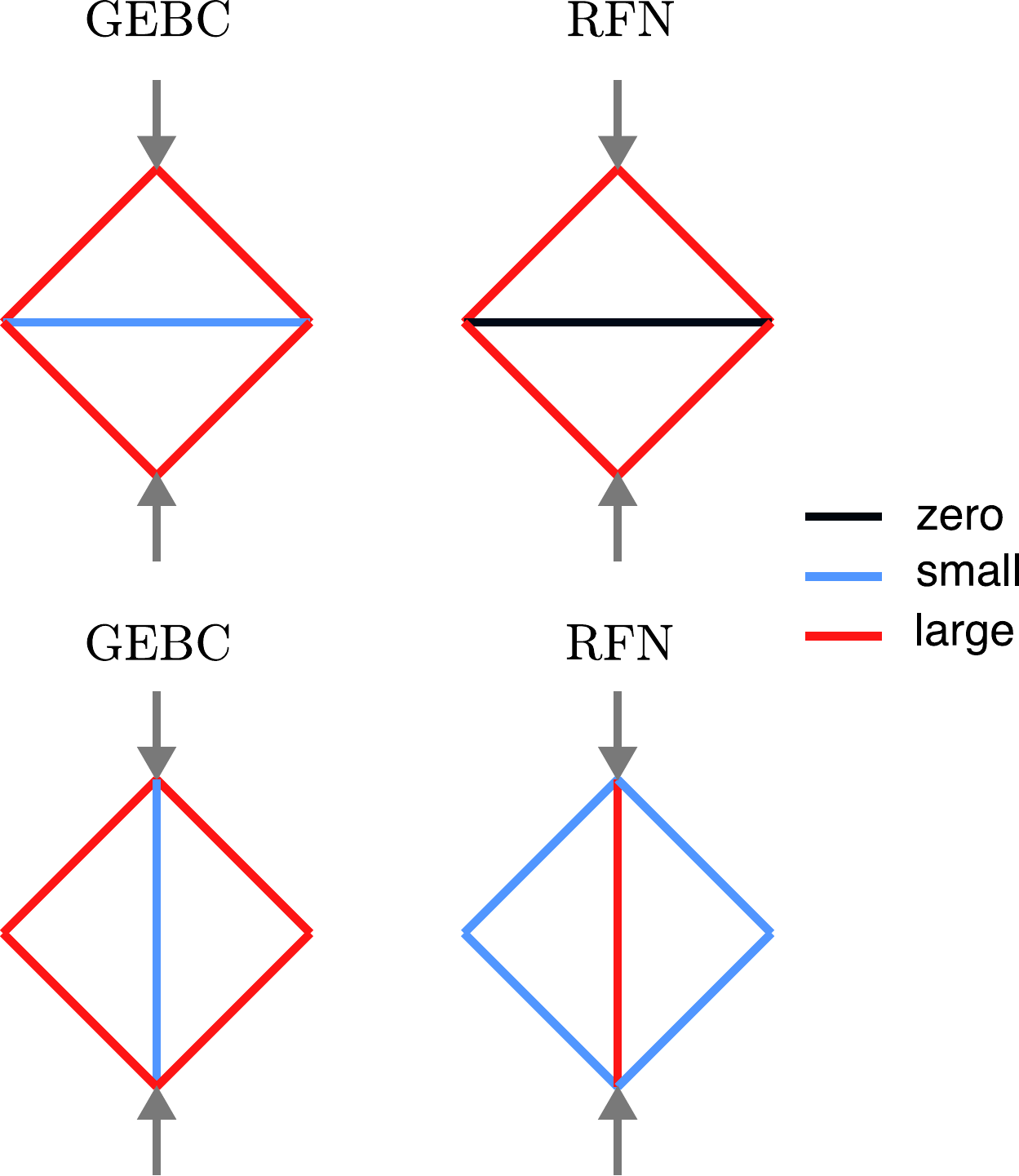}
\caption{Random-fused network versus geodesic edge betweenness centrality at the scale of few edges. When we load the top and bottom nodes, the GEBC
distribution is the same in the top and bottom configurations; by contrast, the RFN current distribution is different in these two configurations. }\label{fig:S5}
\end{figure}

In Fig.~\ref{fig:S5}, we show a schematic that highlights the fundamental difference between the RFN model and GEBC. Specifically, the latter is independent of the loading direction, whereas the former depends on it. In the simple example in this schematic, when we load the top and bottom nodes (see the arrows), we observe that the stresses in each beam change if we rotate the configuration by $90$ degrees. This is captured successfully by the RFN, which identifies the fact that different current values flow in each edge. Note, however, that the RFN model does not systematically predict the correct stress distribution: the central horizontal edge (in the top configuration) has a null current, whereas this beam is stressed if we load the corresponding beam structure. By contrast, the GEBC values do not change when we rotate the structure. Therefore, GEBC is unable to capture physical properties at the scale of a few edges. Nevertheless, as these effects effectively average over multiple directions in our disordered lattices, GEBC and the RFN measure similar values on average, and the GEBC-based test performs well at the scale of our samples.

\end{document}